\documentstyle[aps,epsf,twocolumn]{revtex}
\skip\footins 15.25pt plus 4pt minus 2pt
\def\footnoterule{\kern-5.25pt\hrule width.5in\kern3.6pt}
\renewcommand{\mathrm}[1]{{\rm #1}}

\begin{document}
\draft
%%%%%%%%%%%%%%%%%% TITLE %%%%%%%%%%%%%%

\title{\large\bf Bridged-assisted electron transfer.
Random matrix theory approach.
\vskip 0.5cm
}

\vspace{24pt}
%%%%%%%%%%%%%%%%%%%%%%% AUTHOR(S)
%%%%%%%%%%%%%%%%%%%%%%%

\author{ 
	Ewa {\sc Gudowska-Nowak}\ ${}^{1,3}$
%	\footnote[3]{E-mail:{\tt gudowska@pc.chemie.th-darmstadt.de}}
	G\'abor {\sc Papp}\ ${}^{2}$
%	\footnote[4]{E-mail:{\tt G.Papp@gsi.de}}
	and J\"urgen {\sc Brickmann}\ ${}^{1}$}
%	\footnote[5]{E-mail:{\tt brick@pc.chemie.th-darmstadt.de}} 

\vspace{8pt}
%%%%%%%%%%%%%%%%%%%%%%% ADDRESS
%%%%%%%%%%%%%%%%%%%%%%%
\address{
${ }^{1}$  {\sl Institute for Physical Chemistry
Technische Universit\"at Darmstadt,} \\ 
{\sl Petersenstr. 20,
D-64287, Germany;}\\
${ }^{2}$ {\sl GSI, Plankstr. 1, D-64291 Darmstadt, Germany \&}  \\
	{\sl Institute for Theoretical Physics, E{\"o}tv{\"o}s University, 
	H-1088 Budapest, Hungary;}\\
${ }^{3}$ {\sl Institute of Physics, Jagiellonian University, 30-059
	Krak\'ow, Poland.} 
}
  
%%%%%%%%%%%%%%%%%%%%%% DATE
%%%%%%%%%%%%%%%%%%%%%%%%%%%
\date{\today}
\maketitle

\vspace{6pt}

\begin{abstract}

%%%%%%%%%%%%%%%%%%%%%% ABSTRACT
%%%%%%%%%%%%%%%%%%%%%% 

We discuss the effective donor/acceptor coupling for a bridged electron
transfer system \cite{NEWTON} with a site-diagonal disorder of bridge
energies. The average spectral properties of the system are discussed by
using the Wegner model (Anderson's type tight-binding Hamiltonian (TBH))
for the electronic part of the problem. Spectral properties of the
system are
discussed using the concept of the functional inverse of the resolvent
(``Blue's function'', \cite{ZEE}) for various limits of noise versus
site-site coupling ratio.
\end{abstract}
\addtocounter{page}{-1}

\newcommand{\gm}{\gamma}
\newcommand{\ee}{\epsilon}
\renewcommand{\th}{\theta}
\newcommand{\Sg}{\Sigma}
\newcommand{\dl}{\delta}
\newcommand{\SSg}{\tilde{\Sigma}}
\newcommand{\eq}{\begin{equation}}
\newcommand{\eqx}{\end{equation}}
\newcommand{\eqn}{\begin{eqnarray}}
\newcommand{\eqnx}{\end{eqnarray}}
\newcommand{\ben}{\begin{eqnarray}}
\newcommand{\een}{\end{eqnarray}}
\newcommand{\f}[2]{\frac{#1}{#2}}
\newcommand{\ra}{\rangle}
\newcommand{\la}{\langle}
\newcommand{\bra}[1]{\la #1|}
\newcommand{\ket}[1]{| #1\ra}
\newcommand{\GG}{{\cal G}}
\renewcommand{\AA}{{\cal A}}
\newcommand{\GR}{G(\ee)}
\newcommand{\MM}{{\cal M}}
\newcommand{\BB}{{\cal B}}
\newcommand{\ZZ}{{\cal Z}}
\newcommand{\DD}{{\cal D}}
\newcommand{\HH}{{\cal H}}
\newcommand{\RR}{{\cal R}}
\newcommand{\arr}[4]{
\left(\begin{array}{cc}
#1&#2\\
#3&#4
\end{array}\right)
}
\newcommand{\arrd}[3]{
\left(\begin{array}{ccc}
#1&0&0\\
0&#2&0\\
0&0&#3
\end{array}\right)
}
\newcommand{\tr}{\mbox{\rm tr}\,}
\newcommand{\One}{\mbox{\bf 1}}
\newcommand{\pauli}{\sg_2}
\newcommand{\cor}[1]{<{#1}>}
\newcommand{\cf}{{\it cf.}}
\newcommand{\ie}{{\it i.e.}}
\newcommand{\br}[1]{\overline{#1}}
\newcommand{\phib}{\br{\phi}}
\newcommand{\psib}{\br{\psi}}
\newcommand{\zb}{\br{z}}
\newcommand{\qb}{\br{q}}
\newcommand{\lm}{\lambda}
\newcommand{\ksi}{\xi}

\newcommand{\Gb}{\br{G}}
\newcommand{\Vb}{\br{V}}
\newcommand{\Gm}{G_{q\br{q}}}
\newcommand{\Vm}{V_{q\br{q}}}

\newcommand{\ggd}[2]{\GG_{#1}\otimes\GG^T_{#2}\Gamma}

\section{Introduction}

\noindent 
Electron transfer (ET) processes play a fundamental role in
chemistry~\cite{ULSTRUP} and biology~\cite{MARCUS} with the range of
their abundance ranging from photosynthesis and oxidative
phosphorylation to molecular electronic design. In numerous biological
examples of ET reaction, a single electron is tunneling in an
inhomogeneous medium over large distances of several angstroms. The
intervening medium can be either a protein backbone or a sequence of
cofactors embedded in a protein matrix.  Due to a large separation
between the donor and acceptor, direct electronic coupling between the
chromophores is negligible, rendering thus the question on the effect of
medium on enhancement of the electronic coupling~\cite{ONUCHIC}. The
typical suggestion is that the ET is mediated through the medium which
is acting as a bridge which provides virtual states for the tunneling
electron~\cite{CONNEL}. In such a superexchange mechanism, effective
electronic coupling between the initial and final states depends on the
structure and flexibility of the bridging system.

\noindent 
In this paper, we adopt a standard "effective two-level system" approach
towards ET problem~\cite{ONUCHIC,LARSSON,KARPLUS}. The technique has
been elaborated in the past years by formalizing ET in the energy domain
by use of diagonalization procedures which lead to transition amplitudes
expressed in terms of Green function~\cite{KARPLUS,RATNER,}. To include
possible fluctuations in bridge energies, we assume that the bridge
Hamiltonian can be "sampled" from the class of random Hamiltonians.  The
underlying argument is that inhomogeneous, protein medium can affect the
site energies of the bridge and influence the kinetic ET rate.  Our
studies are performed in the limit of $N\rightarrow\infty$ where $N$
stands for number of bridge orbitals. For such an extended system, we
study effect of the noise on density of bridge electronic states.

\noindent 
The analysis is performed within the framework of the random-matrix
theory~\cite{MEHTA} which turned out to be quite general and a powerful
phenomenological approach to a description of various phenomena such as
quantum chaos~\cite{GUTZWILLER}, complex nuclei~\cite{PORTER}, chaotic
scattering~\cite{MAHAUX} and mesoscopic physics~\cite{HAAKE}. Aspects of
vastly different physical situations such as electron localization
phenomena in disordered conductors and semiconductors~\cite{IIDA},
disordered quantum wires~\cite{MIRLIN} and quantum Hall
effect~\cite{WEIDE} can be described in the language of the random
matrix theory. In all the realms mentioned above, the Hamiltonian of the
system is rather intricate to be handled or simply unknown. In such
cases the integration of the exact equations is replaced by the study of
the joint distribution function of the matrix elements of the
Hamiltonian $P({\bf H})$.

\noindent 
The natural way of addressing the problems of randomness coupled to
various sources is to use technique of the ``free random
variables''~\cite{ZEE,VOICULESCU,SPEICHER,BREZIN,JANIK}. This method
provides an elegant way of ``linearizing'' the process of determining
the average eigenvalue distributions for convolutions representing an
analogue of the logarithm of Fourier transformation of the usual
convolutions.  Recently, the generalization of the ``addition law'' for
hermitean random matrices to the non-hermitean case has been
derived~\cite{JANIK,ZEENEW} and applied by us~\cite{US} to study
properties of a dissipative two-level system.  Similar diagrammatic
approach has been used by other authors~\cite{WANG} to investigate
spectral properties of the Fokker-Planck operator that describes
particles diffusing in a quenched random velocity field.

\noindent 
The paper is organized as follows. In Sections 2 and 3 we discuss
briefly the model Hamiltonian reporting results known from the
literature. Spectral properties of the bridge Hamiltonian are further
estimated in the large $N$ limit. In Section 4 a model of
disorder superimposed on a tight binding Hamiltonian of Sec.2 is presented.
 By use
of the concept of "free variables", we estimate Green function of the
disordered system. Average properties of the system can be further
inferred by studying the structure of the distribution of eigenvalues of
the bridge Hamiltonian which follows derivation of the Green function
for the model.

\section{Partitioning technique, effective coupling and 
Green function of the bridge.}

\noindent 
The transferring system is assumed to have an electronic part described
by a tight-binding (H\"uckel) Hamiltonian \cite{ONUCHIC,KARPLUS,GOLDMAN}:

\ben
H=\ee (\ket D \bra D +\ket A \bra A)+ H_{coupl} + H_{bridge}
\label{ham1}
\een
where $\ee$ stands for the energy of both donor and acceptor states
$\ket D, \ket A$ (for simplicity they are taken here to be equal) and
the coupling $H_{coupl}$ and bridge  $H_{bridge}$ Hamiltonians are given
by

\ben
H_{coupl}=\sum_i^N \big(\beta_{Di}\ket D \bra {b_i} + \beta_{Ai}\ket A\bra
	{b_i}\big) + {\it hc} 
\een
and
\ben
H_{bridge}=\sum_i^N \ee_i \ket {b_i} \bra {b_i} + \sum_{i\neq j}
	\beta_{ij}\ket {b_i}\bra {b_j}   
\label{hbridge}
\een
with $\beta_{ij}$ being a symmetric site-diagonal matrix.

\noindent 
The effective Hamiltonian is obtained by the standard Wigner--Weisskopf
reduction~\cite{LARSSON,WIGNER,LOWDIN} by partition.  In this method the
total Hilbert space of the problem is divided into two subspaces
(spanned by donor and acceptor states and bridge states, respectively)
and the Hamiltonian is integrated over the elements of one of them
(here, the bridge states), eventually mapping the eigenvalue problem of
a high dimension onto a lower one. The time-independent Schr\"odinger
equation for the partition scheme is \cite{LOWDIN}

\ben
({\bf H} -E {\bf I}){\bf c}={\bf M}{\bf c}=0
\een
with matrix ${\bf M}$  given by

\ben
{\bf M}=\arr{{\bf M}_{aa}}{{\bf M}_{ab}}{{\bf M}_{ba}}{{\bf M}_{bb}} \,.
\een
By eliminating subspace ``b'' (spanned by intervening bridge states
$\ket {b_i}$), the effective Schr\"odinger equation becomes

\ben
({\bf M}_{aa}- {\bf M}_{ab}{\bf M}^{-1}_{bb}{\bf M}_{ba}){\bf c}_a =
	\hat{{\bf M}}_{aa}{\bf c}_a =0 
\een
so that the reduced effective Hamiltonian of the system is the 
$2\times 2$ matrix 

\ben
H_{eff}=\hat{{\bf M}}_{aa} + E {\bf I}_a
\een
with energy $E$  set up (in zeroth order) to $\ee$, {\it i.e.} to
donor/acceptor energy. The formalism yields the effective coupling given
by \cite{KARPLUS,GOLDMAN} 
\ben
H_{DA}=-\sum_{ij} \beta_{Di} [{\bf M}^{-1}_{bb}]_{ij}\beta_{jA}
\label{coupl}
\een
where bracketed expression stands for the  Green function of the bridge

\ben
{\bf G}(\ee)=({\bf H}_{bridge} -\ee)^{-1}={\bf M}^{-1}_{bb}
\label{gbridge}
\een
with ${\bf H}_{bridge}$ given by the TBM,
\ben
H^{mn}_{bridge}=\ee_b \delta_{mn} + \beta (\delta_{m, n+1}
	+\delta_{m,n-1}) \,.
\label{hmn}
\een
The Hamiltonian above describes a system with $N$ localized sites, each
with energy $\ee_b$ and a hopping between the
neighboring sites with hopping parameter $\beta$.

{}From the spectral representation of the Green's function (\ref{gbridge})

\ben
G_{mn}\equiv \sum_i\frac{u^i_m u^i_n}{\ee-\ee_i}
\label{spec}
\een
with eigenvalues
\ben
\ee_i=\ee_b + 2\beta\cos(\frac{\pi i}{N+1})
\een
and  spectral coefficients
\ben
u^i_m=(\frac{2}{N+1})^{1/2}\sin(\frac{\pi m i}{N+1}) ,
\een
it follows, that 
\ben
Tr{G(\lambda)}=\frac{1}{N}\sum_i\frac{1}{\lambda-\ee_i}\rightarrow \int d
	\ee \frac{\nu (\ee)}{\lambda-\ee} 
\een
where in the limit of $N\rightarrow\infty$ we have replaced the sum
in~(\ref{gbridge}) by an integral with the level density $\nu(\ee)$
as a weight. The localization length $\gamma^{-1}$ measuring the decay
of an  eigenvector over the chain composed of $N$ units can be deduced
from~eq.(\ref{spec})~\cite{ZIMAN}

\ben
\gamma_{\nu}=\frac{1}{N} \sum_{\mu\neq\nu} 
	\log |\ee_{\nu}-\ee_{\mu}|-\log\beta
\label{length}
\een
and in the limit of an infinite chain reads

\ben
\gamma(\ee_\nu)=\int d\ee\ \nu(\ee) \log|\ee_{\nu}-\ee|-\log\beta
\label{length1}
\een
bringing the dependence on the density $\nu(\ee)$.

\noindent 
By changing to complex variable, the density of eigenvalues can be
conveniently defined in terms of the trace of the resolvent of the
Hamiltonian 
\ben
G(\ee)=\frac{1}{N} \ {\rm Tr \,} 
	\left< \frac{1}{\ee-H_{bridge}}\right> \,.
\label{reso}
\een
The density of states for $H_{bridge}$ is then given by
\ben
\nu(\ee)=\frac{1}{N}\ {\rm Tr \,} \left< \delta(\ee-H_{bridge})\right>=
	-\frac{1}{\pi} Im G(\ee+i\lambda)
\een
and follows from the discontinuities of the resolvent (\ref{reso}) along
the $\ee$-axis.

\section{Kinetic rate.}

\noindent
The kinetic rate for the electron transfer mediated by the bridge can be
evaluated according to the definition \cite{NEWTON,ULSTRUP,ONUCHIC}
\ben
k(t)=\frac{d}{dt} |\left<\phi_f|\psi_I(t)\right>|^2
\label{rate}
\een
where $|\phi_f\rangle$ is a final state and
\ben
 \psi_I(t)=e^{\frac{it}{\hbar} H_0}e^{-\frac{it}{\hbar} H}\psi_i
\een
with $\psi_i$ standing for the exact eigenstate of the Hamiltonian
(\ref{ham1}).  The latter can be decomposed into $H=H_0+H_1$ in which
$H_0$ describes diagonal part of the total Hamiltonian and $H_1$ defines
interaction inducing charge transfer between the donor and
acceptor~\cite{ULSTRUP}. The exact eigenstates of $H$ can be represented
as
\ben
\psi_i=\phi_i + \frac{1}{\ee_i-H_0+i\lambda}H_1\psi_i\nonumber \\
=\phi_i + \frac{1}{\ee_i-H+i\lambda}H_1\phi_i
\een
where $\phi_i$ is the initial state of the system and
\ben
T=H_1+H_1\frac{1}{\ee_i-H+i\lambda}H_1
\een 
stands for the transition operator which to the lowest order of approximation 
is given by
\ben
T=H_1+H_1\frac{1}{\ee_i-H_0+i\lambda}H_1 \,.
\een
By use of the transition operator, the kinetic rate (\ref{rate}) can be
rephrased in the form 
\ben
k=\frac{2\pi}{\hbar}|\left<\phi_f|T|\phi_i\right>|^2\delta(\ee_f-\ee_i)
	\nonumber \\ 
 =\frac{2\pi}{\hbar}|\left<\phi_f|H_1|\psi_i\right>|^2\delta(\ee_f-\ee_i)
 \,.
\een
By assuming that the vibrational relaxation in the molecular system is much
 faster than the electron transfer, the above equation can be expressed as
\ben
k=\frac{2\pi}{\hbar}\sum_{v'}\sum_v P_{iv}\left|\left<\Psi_{fv'}|T|
|\Psi_{iv}\right>\right|^2\delta(\ee_{fv'}-\ee_{iv})
\label{rate1}
\een
where $(if)$ and $(v,v')$ stand for electronic and vibronic states,
respectively and $P_{iv}$ is Boltzmann weight factor. In the adiabatic
approximation, the wavefunctions of the system can be written as
products of electronic and vibrational wavefunctions
\ben
\Psi_{iv}=\Phi_i\Theta_{iv} \,.
\een
Substitution of the above approximation to the evaluation of matrix
elements of $T$ yields
\ben
\label{trans}
\left<\psi_{fv''}|T|\psi_{iv}\right> 
	\hspace*{-15mm}&\hspace*{15mm}=\hspace*{-15mm}&\hspace*{15mm}
	\left<\Theta_{fv''}|H_{1fi}|\Theta_{iv}\right>
	\\ 
  &+& \sum_{m,v'}\frac{%
	\left<\Theta_{fv''}|H_{1fm}|\Theta_{mv'}\right>
	\left<\Theta_{mv'}|V_{mi}|\Theta_{iv}\right>}%
	{\ee_{iv}-\ee_{mv'}+i\lambda}. \nonumber
\een

In the above formula, elements $H_{1fm}$ represent electronic matrix
elements between states $\Phi_f$ and $\Phi_m$. {\it Condon}
approximation~\cite{ULSTRUP,MARCUS,ONUCHIC} allows for factorization of
these terms from the {\it Franck-Condon} factors
$\left<\Theta_{fv''}|\Theta_{iv}\right> $, and in the case when the
electronic gap between the donor and acceptor states is larger than
vibrational energies in the system, leads to the ET rate
\ben
k=\frac{2\pi}{\hbar} |T_{fi}|^2\sum_{v',v} P_{iv}\left|\left<\Theta_{fv'}
|\Theta_{iv}\right>\right|^2\delta(\ee_{fv'}-\ee_{iv}).
\label{rate2}
\een
The rate constant can be thus written as a product of electronic part
and nuclear part, averaged over vibrational states. Throughout the paper
we will focus on electronic degrees of freedom and estimation of the
transition matrix~(\ref{trans}).\\ 
Note, that by neglecting a direct electronic coupling between the donor and 
acceptor and interpreting 
\ben
\left<\Phi_{f}|H_1|\phi_{N}\right>=\left<\phi_{1}|H_1|\Phi_{i}\right>=\beta_1
\een
the formula for the transition matrix~(\ref{trans}) is identical with the
the effective coupling~(\ref{coupl}) and for the chain of $N$ elements of 
the bridge reduces to~\cite{KARPLUS}
\ben
T_{fi}\equiv H_{DA}=(-1)^N\frac{{\beta_1}^2\beta^{N-1}}{(\ee_b-\ee)^N}=
	-\beta_1^2G_{1N}
\label{transfer}
\een
where the element $G_{1N}$ of the Green's function is
\ben
G_{1N}=\frac{(-1)^{N+1}\beta^{N-1}}{det H_{bridge}}
\label{func}
\een
with $det$ standing for the determinant of the bridge Hamiltonian (\ref{hmn}):
\ben
det_N = 
	\frac{\beta^N[(\alpha+\sqrt{\alpha^2-1})^{N+1}-
	(\alpha-\sqrt{\alpha^2-1})^{N+1}]}%
	{2\sqrt{\alpha^2-1}}
\een
with
\ben
\alpha=\frac{\ee-\ee_b}{2\beta}\,.
\een
In the $N\to\infty$ limit $det$ approaches $\beta^N$ for $|\alpha|<1$
and $\infty$ otherwise, thus the transfer element $G_{1N}$ is $1/\beta$
for a small donor-bridge energy splitting ($|\alpha|<1$), and 0 for
tunneling energies outside the band of the width $2\beta$ (the latter
corresponds to the Bloch-like extended states of the bridge
\cite{KARPLUS}).

Localization constant introduced in the former paragraph is related to
the decay of the transfer matrix~\cite{ZIMAN}
\ben
\gamma  = - \lim_{N\rightarrow\infty}\frac{1}{N}\ln|G_{1,N}|
\label{gamma}
\een
and, in the case of disordered bridge, would require using the density
function $\nu(\ee)$ to estimate the average~(\ref{length1}). In the
deterministic case (no site diagonal `disorder') one has
\ben
  \gamma = \left\{ \begin{array}{cc}
	0 & |\alpha| < 1 \cr
    \mbox{max} \log{|\alpha\pm\sqrt{\alpha^2-1}|}
	& |\alpha| > 1 \end{array} \right. \,.
\een
so that the states with energies lying within the band $(-2\beta, 2\beta)$
are localized over the infinite range.

\section{Transition through the disordered chain. Random Matrix models.}

\noindent 
As it stands, the bridge Hamiltonian ({\sl tight-binding Hamiltonian}
TBH, eq.~(\ref{hbridge})), can be generalized, as in the case of the
Anderson model, to include disorder of the bridge chain. In what
follows, we will adopt the model of the site diagonal disorder which is
equivalent to the ones studied by {\sl Wegner}~\cite{WEGNER} and {\sl
Neu and Speicher}~\cite{SPEICHER,NEU}.  The idea of Wegner was to
generalize the Anderson model by putting $n$ electronic states at each
site of the $d$--dimensional lattice and describing the disorder by
Gaussian random matrices in the electronic states. For $n=1$ Wegner's
model reduces to the usual unsolvable Anderson model and becomes exactly
solvable for $n\rightarrow\infty$\footnote{The method described in that
paper also applies~\cite{NEU,NELSON} to the original $n=1$ Anderson
model and then it is equivalent ~\cite{NEU,NELSON} to the
CPA (``coherent potential'') approximation~\protect\cite{ZIMAN}, which
works remarkably well. That allows us to extrapolate the results to finite
$n$ configurations.}. Further generalization of Wegner's
model has been discussed by {\sl Neu and Speicher} who have used a
mathematical concept of ``freeness'' to use a more general ensembles of
random matrices allowing arbitrarily distributed disorder. ``Freeness'',
introduced in mathematical literature by {\sl Voiculescu, Pastur and
Speicher}~\cite{VOICULESCU,SPEICHER,PASTUR} has been also popularized
recently in physical applications by {\sl Brezin, Zee, Janik et
al.}~\cite{ZEE,BREZIN,JANIK}.

\noindent 
The formalism can be translated to describe spectral properties of a
Hamiltonian of the form 
\eq
H=H^D + H^R
\eqx
where $H^D$ is deterministic, and $H^R$ random part of the operator. By
assuming that $H^D$ and $H^R$ are free with respect to average over the
disorder, {\sl Voiculescu, Pastur, Neu and Speicher} have shown that the
diagonal part of the one-particle Green function associated with the
total Hamiltonian $H$ satisfies equation

\eq
G(\ee)=G^D[\ee-\Sigma\big(G(\ee)\big)] \,.
\label{voic}
\eqx
where the argument of $G^D$ is $\ee-\Sigma$ with $\Sigma$ being nothing
but the self energy determined by
\eq
G^R=\frac{1}{\ee-\Sigma(G^R)}
\eqx
where
\eq
G^R=\frac{1}{N}Tr\ \left<\frac{1}{\ee-H^R}\right>\,.
\eqx 
\noindent 
The same result has been rederived by {\sl Zee}~\cite{ZEE} who through
his diagrammatic analysis, introduced the "Blue's function" that is just
the functional inverse of the resolvent
\eq
B[G(\ee)]=\ee
\label{inverse}
\eqx
and satisfies the additivity law
\eq
B^{D+R}(\ee)=B^D(\ee)+B^R(\ee)-\f{1}{\ee}\,.
\label{sumblue}
\eqx
Both equations~(\ref{voic}) and~(\ref{sumblue}) coincide if one identifies 
$B(\ee)=\Sigma(\ee)+\ee^{-1}$.

\noindent 
Our further analysis is based on the assumption that nodal energies of
the bridge are randomly distributed with off-diagonal elements of matrix
$H_{bridge}$ being constant.  In the limit of large $N$, deterministic
resolvent of the bridge Hamiltonian yields
\ben
G^D(\ee) &=& 
	\frac{1}{N}\sum_{k=1}^{N} \frac{1}{\ee-2\beta\cos\frac{\pi k}{N+1}}
	   \nonumber\\
	&\stackrel{\mbox{\tiny $N\!\!\!\to\!\!\infty$}}{\longrightarrow}&
%	(N\rightarrow\infty) \,\,&=&
	\int^1_0\frac{dx}{\ee-2\beta\cos\pi x} 
%	\nonumber \\
     =\frac{1}{\sqrt{\ee^2-4\beta^2}} %, \,\, for\, \ee^2>4\beta^2
\label{limit}
\een
(note, that the average $\ee_b=0$ has been set up to 0) with the
deterministic "Blue's function"
\eq
B^D=\sqrt{\frac{1}{\ee^2}+4\beta^2} \,.
\eqx
Evaluation of the deterministic  $G_{1N}$ element for this case leads,
in the limit of $({\ee_b-\ee})/2\beta > 1$, to the usual {\sl McConnel}
result~\cite{CONNEL}
\ben
\lim_{N\rightarrow \infty}G_{1N}=0 \,.
\een
Qualitative estimate of the transfer matrix for the ``random plus
deterministic'' case follows now, according to the formula~(\ref{func}),
determination of the inverse determinant for the full (diagonal random
plus TBH deterministic) bridge Hamiltonian:
\ben
G_{1N}=\left<\frac{(-1)^{N+1}\beta^{N-1}}{det(H^D+H^R)}\right> \,.
\een 
For $\beta$ ``large'' $det$ is dominated by $\beta$, giving
$det=\beta^N$; for small $\beta$ the determinant is dominated by the
randomness of the diagonal elements in the bridge Hamiltonian.  Thus
naively, in the limiting case of ``large'' $\beta$ (and for an infinite
bridge) $\beta$ in the numerator is expected to be negligible compared
to the $det$ leading to $G_{1N}$ proportional to $1/\beta$. More careful
analysis (see below) shows, that randomness always increases $\gamma$,
leading to the faster decay of the transfer matrix $G_{1N}$ over the
distance.

If randomness of $\ee_b$ is chosen to be sampled with the semi-circle 
law\footnote{We note that choosing other randomness does not change our
results qualitatively.}
\eq
\varrho(\ee_b)=\frac{1}{2\pi\sigma^2}\sqrt{4\sigma^2-\ee_b^2}
\eqx
which is equivalent to a random matrix model \cite{MEHTA,PORTER,WEGNER}
of the noisy contribution $H^R$ to the bridge Hamiltonian, Green's
function associated with $H^R$ takes the form (see eq.~(\ref{voic}))

\eq
G^R(\ee)=\frac{1}{\ee-\sigma^2 G^R(\ee)}=
	\frac{\ee\mp \sqrt{\ee^2-4\sigma^2}}{2\sigma^2}\,.
\eqx
The functional inverse of $G^R$ is given by Blue's function
\eq
B^R(\ee)=\sigma^2 \ee + \frac{1}{\ee}
\eqx
and, after using the addition law~(\ref{sumblue}) leads to the following
equation for the  Green's function of the system
\eq
\sigma^4 G^4- 2\ee\sigma^2 G^3 + (\ee^2-4\beta^2)G^2-1 = 0 \,.
\label{Pa}
\eqx
The end-points of the spectra may be calculated from the equation
\ben
\frac{dG}{d\ee}|_{\ee=a}=\infty
\een
or, equivalently from the discriminant of the above equation. By
introducing the rescaled variables
\ben
  \frac \ee{\beta} \to \ee,  \qquad \frac{\sigma}{\beta} \to \sigma,
\qquad \beta G\to G
\label{scal}
\een
the discriminant (end-point condition) for the Pastur equation~(\ref{Pa}) is
\ben
  4 \ee_e^6+\!(\sigma^4\!\!-\!48) \ee_e^4+\!16 (12\!-\!5\sigma^2) \ee_e^2
        \!-\!16 (4\!+\!\sigma^4)^2\!= 0 \,.
\een
In the limit $\sigma\rightarrow 0$ that reduces to $(\ee_e^2-4)^3=0$, thus
the support of the spectrum is {\em one} interval, with end-points $\pm
2$, while in the limit $\sigma\rightarrow\infty$ ($\beta\rightarrow0$)
one gets once again {\em one} interval with end-points $\ee=\pm
2\sigma$. Generally, the discriminant equation has only one pair of real
solution for any value of $\sigma$.

\begin{figure}[t]
\centerline{\epsfysize=5truecm \epsfxsize=7truecm \epsfbox{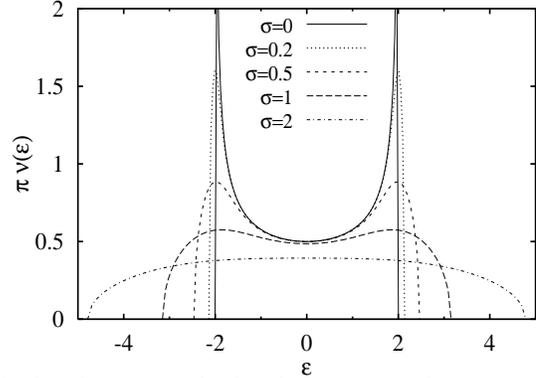}}
\caption{Eigenvalue distribution for the TBH of the bridge for different
values of the noise/deterministic ratio, $\sigma$.}
\label{fig-spec}
\end{figure}

The spectra for different values of $\sigma$ is shown in
Fig.~\ref{fig-spec}. `Deterministic' spectrum is divergent close to the
the endpoints of the band. For an increasing value of the noise intensity
$\sigma$, the distribution $\nu(\ee)$ flattens and, eventually
changes from the bimodal function with two peaks located at the ends of
the support to the unimodal distribution with a broad hump around the
center of the support.  The curvature around zero changes sign for a
critical value of the noise intensity which can be found from~(\ref{Pa})
analyzed around $\ee=0$. With the scaling~(\ref{scal}) one gets then the
critical value of $\sigma_c $ in units of $\beta$
\ben
\sigma_c= 12^{1/4}= 1.8612.
\een 
Note that from the characteristics of $\nu(\ee)$ it follows, that the
localization length $\gamma^{-1}$ has a discontinuity at the end-points
of the spectra for $\sigma=0$. For energies within the band, $\gamma$ is
zero and switches to a finite value after the condition $|\alpha|>1$ is
reached. The noise in some sense softens the abrupt change of the
$\gamma$ at the endpoints of the support (\cf Fig.2) and shortens the
localization length within the band.

\begin{figure}[t]
\centerline{\epsfysize=5truecm \epsfbox{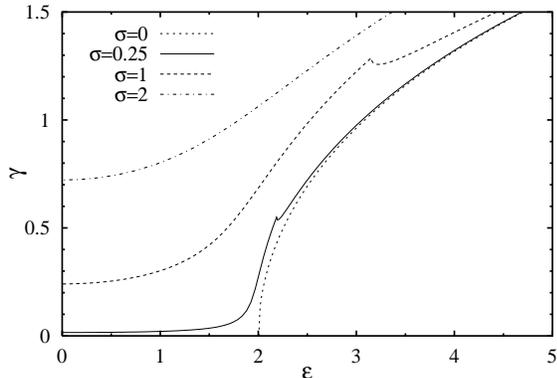}}
\caption{Decay coefficient $\gamma$ as a function of energy for various
intensities of the noise. $\beta$ has been set up to 1. The completely
deterministic system ($\sigma=0$) has $\gamma=0$ for $|\ee|<2$.}
\label{fig-gamma}
\end{figure}

The ``deterministic plus random'' $G_{1N}$ can be evaluated (in the
limit of large $N$) following Neu and Speicher~\cite{SPEICHER},
\ben
  G_{1N}(\ee) = G^D_{1N} (\ee-\Sigma[G(\ee)])
\label{offdiag}
\een
\ie, the same {\sl Pastur} equation~(\ref{voic}) holds also for the
non-diagonal elements of the total Green's function. The general
analysis of Eq~(\ref{offdiag}) shows, that the solution $\gamma=0$ can
not be achieved any more for any combination of $\ee_b$ and
$\beta$. However for ``small'' disorder the states with energies
$({\ee_b-\ee})/2\beta\le 1$ are stable with small decay coefficient
(see Fig.~\ref{fig-gamma}). The most stable energies are sampled around
$\ee=0$ with decay coefficients
\ben
 \gamma = \log{\frac 1{\sqrt{2}} \left( 
	  \sqrt{\sqrt{1\!\!+\!\!\sigma^4/2}-1} + 
	  \sqrt{\sqrt{1\!\!+\!\!\sigma^4/2}+1}
	\right)}
\een
yielding $\gamma=0$ only for vanishing randomness ($\sigma=0$).

In  Fig.~\ref{fig-gamma} one may note the cusp-like discontinuities
around the end-points 
$\ee_e$ of the spectrum. Their appearance may be understood analytically
in the large $\sigma$ limit, where by evaluating the resolvent $G$ we
neglect the deterministic part of the Hamiltonian. The resolvent then
reads  
\ben
  G^R(\ee)=\frac{\ee\pm\sqrt{\ee^2-4\sigma^2}}{2\sigma^2}
\een
with end-points $\ee_e=\pm2\sigma$. Evaluating~(\ref{offdiag}) and
inserting it into~(\ref{gamma}) the expansion around the end-points
$\ee_e$ for $\ee>\ee_e$ leads to
\ben
  \gamma \approx \log{\frac{\sqrt{\sigma^2-4}}2} -
	\frac{\sqrt{\sigma}}{\sqrt{\sigma^2-4}} (\ee-\ee_e)^{1/2} + ...
\een
showing the observed ``cusp'' around the end-point.

\section{Conclusions}
%\vskip 2cm

\noindent 
In this paper we have discussed briefly properties of a bridged
electron transfer affected by site diagonal disorder in the bridge
Hamiltonian. Instead of using the standard language of the Green's function
formalism, we have chosen to work with the functional inverse of the 
Green's function (``Blue's function''). The method stems from the powerful
mathematical concept of free random variables~\cite{VOICULESCU,SPEICHER}
which has been shown to be 
an elegant tool in  various applications of the random matrix
theory~\cite{BREZIN,JANIK,ZEENEW,US,WANG,NELSON1}.

Our model refers to the situation when the bridge is long
enough to be considered infinite. Site diagonal disorder is assumed in
the form of a random matrix model resulting in placing at each site of
the bridging chain a random matrix with a semi-circular distribution of
energies ({\sl Wegner} model, \cite{WEGNER}). That would correspond to a
situation where the bridge energies are n-electron functions.  Presence
of noise extends and flattens the spectrum of the TBH Hamiltonian. For
donor/acceptor energies sampled from the center of the TBH band, the
noise increases the inverse localization length $\gamma$ and leads to a
fast decay of the
electronic coupling with the distance measured in bridge units. The
kinetic rate becomes thus vastly reduced by the noise (note that our
discussion relates only to the electronic part of the ET rate).  The
distribution of TBH energies has been shown to posses a noise-induced
characteristics which depends on the value of the critical noise
intensity $\sigma_c$.  As expected \cite{WEGNER,PASTUR,PARDY}, the
diagonal disorder localizes the eigenfunctions of the TBH Hamiltonian
resulting in the reduction of the electronic transfer matrix
$T_{fi}$. Thus the diagonal noise reinforces the exponential decay of
the effective coupling with increasing distance between the donor and
acceptor.\\
The method of Blue's function applied here is also suitable for
nonhermitean ensembles of 
random matrices which are used in quantum theory of
dissipation~\cite{HAAKE,ZEENEW,NELSON}.
The paper discusses so far only the case of the hermitean
ensemble. Extension of the 
formalism to models of dissipative transport will be presented elsewhere.
 
\vskip 12mm

{\bf \noindent  Acknowledgments } \\ 
This project has been supported by the  Deutsche Forschungsgemeinschaft,
Bonn, the Funds der Chemischen Industrie, Frankfurt and by Hungarian
grant FKFP126/97.
\vskip 3mm

\vfill
\eject
%\vglue 2.5cm
{\bf REFERENCES.}

%\bibliographystyle{aip}
%\bibliography{vacref}

\vspace*{-5mm}
\begin{thebibliography}{50}
\vspace*{-13mm}
\bibitem{NEWTON}
T.J. Meyer and M.D. Newton, eds., 
	Chem. Phys. (Special Issue), {\bf 176} (1993). 

\bibitem{ZEE}
A. Zee, 
	Nucl. Phys. {\bf B474} (1996) 726.

\bibitem{ULSTRUP}
J. Ulstrup, 
	{\sl Charge Transfer Processes in Condensed Media}, (Springer
	Verlag, Berlin, 1979) 

\bibitem{MARCUS}
R.A. Marcus and N. Sutin,
	Biochim. Biophys. Acta, {\bf 811} (1985) 265.

\bibitem{ONUCHIC}
S.S. Skourtis and J.N. Onuchic,
        Chem. Phys. Lett. {\bf 209} (1993) 171.

\bibitem{CONNEL}
H.M. McConnel, 
        J.Chem.Phys. {\bf35} (1961) 508.

\bibitem{LARSSON}
S. Larsson, 
	J. Am. Chem. Soc. {\bf 103} (1981) 4034.

\bibitem{KARPLUS}
J.W. Evenson and M. Karplus, 
	J. Chem. Phys. {\bf 96} (1992) 5272.

\bibitem{RATNER}
M. Kemp, A. Roiberg, V. Mujica, T. Wanta  and M.A. Ratner,
        J. Phys. Chem. {\bf 100} (1996) 8349.

\bibitem{MEHTA}
M.L. Mehta, 
	{\sl Random Matrices}, (Academic Press, New York, 1991).

\bibitem{GUTZWILLER}
M.C. Gutzwiller, 
	{\sl Chaos in Classical and Quantum Mechanics}, (Springer Verlag,
	Berlin, 1990).

\bibitem{PORTER}
C.E. Porter, 
	{\sl Statistical Theories of Spectra: Fluctuations}, (Academic
	Press, New York, 1965).

\bibitem{MAHAUX}
C. Mahaux and H.A. Weidenm\"uller, 
	{\sl Shell Model Approach to Nuclear Reactions}, (North Holland,
	Amsterdam, 1969).

\bibitem{HAAKE}
F. Haake, F. Izrailev, N. Lehmann, D. Saher and H.J. Sommers, 
	Z. Phys. {\bf B88} (1992) 359.

\bibitem{IIDA}
S. Iida, H.A. Weidenm\"uller and J.A. Zuk, 
	Ann. Phys. {\bf 200} (1990) 219.

\bibitem{MIRLIN}
A.D. Mirlin, A. M\"uller-Groeling and M.R. Zirnbauer, 
	Ann. Phys. {\bf 236} (1994) 325.

\bibitem{WEIDE}
H.A. Weidenm\"uller and M.R. Zirnbauer, 
	Nucl. Phys. {\bf B305} (1988) 339.

\bibitem{VOICULESCU}
D.V. Voiculescu, 
	Invent. Math. {\bf 104} (1991) 201.

\bibitem{SPEICHER}
R. Speicher,
         Math. Ann. {\bf 298} (1994) 611.

\bibitem{BREZIN}
E. Br\'ezin and  A. Zee, 
	Phys. Rev. {\bf E49} (1994) 2588.

\bibitem{JANIK}
R.A. Janik, M.A. Nowak, G. Papp, J. Wambach and I. Zahed, 
	Phys. Rev. {\bf E55} (1997) 4100;
R.A. Janik, M.A. Nowak, G. Papp and I. Zahed,
	Nucl. Phys. {\bf B501} (1997) 603.

\bibitem{ZEENEW}
J. Feinberg and A. Zee
        {\it Nonhermitean Random Matrix Models}, e-print cond-mat/9703087

\bibitem{US}
E. Gudowska-Nowak, G. Papp and J.Brickmann,
        Chem. Phys. {\bf 220} (1997) 125.

\bibitem{WANG}
J.T. Chalker and Z. J. Wang,
	Phys. Rev. Lett. {\bf 79} (1997) 1797.

\bibitem{GOLDMAN}
C. Goldman, 
        Phys. Rev. {\bf 43A} (1991) 4500.

\bibitem{WIGNER}
E.P. Wigner and V. F. Weisskopf,
         Z. Phys. {\bf 63} (1930) 54.

\bibitem{LOWDIN}
P.O. L\"owdin, 
	J. Math. Phys. {\bf 3} (1962) 969.

\bibitem{ZIMAN}
J.M. Ziman, 
	{\sl Models of disorder}, Cambridge University Press, Cambridge, (1979).

\bibitem{WEGNER}
F. Wegner,
         Phys. Rev. B. {\bf 19} (1979) 783.

\bibitem{NEU}
P. Neu and R. Speicher,
         J. Stat. Phys. {\bf 80} (1995) 1279.

\bibitem{NELSON}
R. Janik, M.A. Nowak, G. Papp and I. Zahed,
         {\it Localization Transitions From Free Random Variables},
         e-print cond-mat/9705098.

\bibitem{PASTUR}
L.A. Pastur,
         Theor. Math. Phys. (USSR) {\bf 10} (1972) 67.

\bibitem{NELSON1}
N. Hatano and D.R. Nelson,
          Phys. Rev. Lett.{\bf 77} (1997) 570.

\bibitem{PARDY}
J.B.  Pendry, 
          Adv. Phys. {\bf 43} (1993) 461.

\end{thebibliography}

\end{document}